\documentclass[aps,prmat,nobibnotes,nofootinbib,superscriptaddress,twocolumn,
]{revtex4-2}

\usepackage{graphicx} 	
\usepackage{dcolumn} 	
\usepackage{bm}   		
\usepackage{amssymb}   	
\usepackage[separate-uncertainty=true,range-units=single,range-phrase=\text{--},retain-explicit-plus=true]{siunitx} 
\usepackage[svgnames]{xcolor}	
\usepackage[version=3]{mhchem}	

	\usepackage{xr} 
	\usepackage[colorlinks=true,citecolor=ForestGreen]{hyperref}
	\usepackage[caption=false]{subfig}
	\usepackage[nameinlink,capitalise]{cleveref}
	\usepackage{lipsum}
	\usepackage{xspace} 
	
	\DeclareSIUnit\unitcell{uc}
	
	\graphicspath{{figs/}}

	\def\Th{T_\mathrm{holder}\xspace}

	\def\TMPMS{T_\mathrm{MPMS}\xspace}
	\def\Ttop{T_\mathrm{top}\xspace}
	
	\def\Tm{T_m\xspace}
	\def\Th{T_\mathrm{hot}\xspace}
	\def\xh{x_\mathrm{hot}\xspace}
	
	\def\Pp{P_\mathrm{2p}\xspace}
	\def\Rp{R_\mathrm{2p}\xspace}
	\def\Rpp{R_\mathrm{4p}\xspace}

	\def\dTtop{\Delta \Ttop\xspace}
	
	\def\dTm{\Delta \Tm\xspace}
	\def\CRO{\ce{Ca2RuO4}\xspace}

\begin{document}
\title{Challenges in extracting nonlinear current-induced phenomena in \CRO}

\author{Giordano Mattoni}
\email{mattoni@scphys.kyoto-u.ac.jp}
\affiliation{Toyota Riken -- Kyoto University Research Center (TRiKUC), Kyoto 606-8501, Japan}
\affiliation{Department of Physics, Graduate School of Science, Kyoto University, Kyoto 606-8502, Japan}

\author{Kazumi Fukushima}
\affiliation{Department of Physics, Graduate School of Science, Kyoto University, Kyoto 606-8502, Japan}

\author{Shingo Yonezawa}
\affiliation{Department of Physics, Graduate School of Science, Kyoto University, Kyoto 606-8502, Japan}
\affiliation{Department of Electronic Science and Engineering, Graduate School of Engineering, Kyoto University, Kyoto 615-8510, Japan}

\author{Fumihiko Nakamura}
\affiliation{Department of Education and Creation Engineering, Kurume Institute of Technology, Fukuoka 830-0052, Japan}

\author{Yoshiteru Maeno}
\affiliation{Toyota Riken -- Kyoto University Research Center (TRiKUC), Kyoto 606-8501, Japan}
\affiliation{Department of Physics, Graduate School of Science, Kyoto University, Kyoto 606-8502, Japan}

\begin{abstract}
	
	An appealing direction to change the properties of strongly correlated materials is to induce non-equilibrium steady states by the application of a direct current.
	%
	While access to these novel states is of high scientific interest, Joule heating due to current flow often constitutes a hurdle to identify non-thermal effects.
	%
	The biggest challenge usually resides in measuring accurately the temperature of a sample subjected to direct current, and to use probes that give direct information of the material.
	In this work, we exploit the simultaneous measurement of electrical transport and magnetisation to probe non-equilibrium steady states in \CRO.
	In order to reveal non-thermal current-induced effects, we employ a simple model of Joule self-heating to remove the effects of heating and discuss the importance of temperature inhomogeneity within the sample.
	%
	Our approach provides a solid basis for investigating current-induced phenomena in highly resistive materials.
	
\end{abstract}

\date{\today}
\maketitle

\section{Introduction}
Direct electric current is a powerful control parameter capable of inducing non-equilibrium steady states in strongly correlated materials \cite{
	kumai1999current,
	myers1999current
}.
Non-equilibrium conditions can be a gateway to peculiar physics that cannot be accessed by other means
\cite{
	fausti2011light,
	stojchevska2014ultrafast,
	mattoni2018light
}.
Despite the large scientific interest, experiments with direct current always entail some extent of Joule heating, which becomes particularly relevant when dealing with highly resistive materials.

The strongly correlated oxide \CRO is a Mott insulator at room temperature that presents a metal--insulator transition (MIT) at about \SI{360}{\kelvin} and a magnetic transition towards an antiferromagnetic state below \SI{108}{\kelvin} \cite{
	nakatsuji1997ca2ruo4,
	cao1999antiferromagnetic
}.
The MIT is characterised by a strong coupling between \CRO lattice and its electronic structure
\cite{
	han2018lattice
}.
Several reports showed that the MIT can be triggered by the flow of electric current, with the suppression of its Mott gap \cite{
	nakamura2013electric,
	okazaki2013current
}.
The occurrence of this current-induced transition has been confirmed by means of neutron scattering \cite{
	bertinshaw2019unique,
	jenni2020evidence
}, electrical transport \cite{
	cirillo2019emergence,
	zhao2019nonequilibrium
}, and also ARPES \cite{
	ootsuki2022metallic,
	curcio2023current
}.
Other reports showed the existence of a rich nanoscale structure at the phase boundary between metallic and insulating regions, possibly induced by the applied voltage rather than the flowing current itself
\cite{
	zhang2019nano,
	vitalone2022nanoscale,
	gauquelin2023pattern
}.
These effects motivate a deeper investigation on the meachanism of electrically-induced states in \CRO.

Measurements with applied current on \CRO are particularly challenging because they involve the application of considerable electrical currents which, at low temperature, lead to the insurgence of a large Joule heating because of the high resistivity of the material.
Heating makes it difficult to evaluate the sample temperature accurately, and in the case of magnetic measurements can also introduce spurious background signals
\cite{
	mattoni2020diamagnetic
}.
Various efforts have been taken to measure accurately \CRO temperature under applied current:
some groups employed thermal imaging \cite{
	okazaki2013current,
	mattoni2020role
},
Fursich \textit{et al.} looked at the shift of the Raman lines \cite{
	fursich2019raman
},
Okazaki \textit{et al.} used a gold nanoparticle to locally assess the sample temperature \cite{
	okazaki2020current
},
Avallone \textit{et al.} employed a nanoscale thermometer patterned right above a tiny \CRO crystal \cite{
	avallone2021universal
}.

In this work, we employ as a ``double probe" magnetic and electrical measurements performed simultaneously to investigate the current-induced state of \CRO.
To do so, we design a special sample holder and thermometer that maximises sample cooling and measures the sample temperature as accurately as possible.
In the presence of non-heating current-induced effects, we expect changes in magnetisation and resistance to have a different dependence on current.
We uncover extensive sample heating that can be explained by the coexistence of a homogeneous and inhomogeneous temperature increase.
We discuss how inhomogeneous temperature profiles may explain remaining nonlinearities and provide a solid basis for probing current-induced effects.

\section{Experimental details}
We performed measurements in a magnetic property measurement system (MPMS) from Quantum Design.
Simultaneous measurements of the sample resistance $R$ and the magnetic moment $m$ were enabled by the custom-made sample holder described in \cref{fig:Setup_MPMS,fig:Setup_copper,fig:Setup_sample}.
The sample holder ensures a large sample cooling, crucial for measurements with applied current, thanks to a large copper strip (dimensions \qtyproduct{200x6x0.4}{\milli\metre}) that constitutes its main body.
The sample holder fits into a standard MPMS plastic straw (diameter \SI{6}{\milli\metre}) that is used for magnetic measurements.
Before sample mounting, the in-plane crystalline axes of \CRO were determined by separate magnetic measurements (\cref{fig:SampleAxes}).
A vertical field $\mu_0 H=\SI{1}{\tesla}$ was applied along the orthorhombic $a$ axis of \CRO, while electrical current is applied along $b$.
This configuration minimises eddy currents in the sample holder and the magnetic background.
Cooling ramps were performed by changing at a rate of \SI{2}{\kelvin\per\minute} the sample-space temperature $\TMPMS$, measured on the outer surface of the copper jacket around the sample space (\cref{fig:Setup_MPMS}).
Helium exchange gas (approximately \SI{1}{\milli\bar} at room temperature) ensures a thermal connection between the sample holder and $\TMPMS$.

Electrical contact to the sample was provided by Au wires (diameter \SI{50}{\micro\metre}) and Ag paint (DuPont 4929N with diethyl succinate, cured at room temperature), which were then linked to thicker copper wires (diameter \SI{0.2}{\milli\metre}).
This configuration leads to a low contact resistance, typically much below \SI{100}{\ohm} at room temperature (\cref{fig:Resistance24p}).
A thin sheet of cigarette paper was used to eletrically insulate the sample from the copper holder, and GE Varnish 7031 was used to glue the elements together and ensure a good thermal contact.
Electrical measurements were performed by sourcing a current to the sample with a Keysight B2912A (voltage compliance \SI{210}{\volt}) and measuring the 2- or 4-probe voltage with a Keithley Electrometer 6514 (input impedance $>\SI{200}{\tera\ohm}$).
The experiments were performed on several \CRO crystals and we here report two representative examples.
Sample \#1 (CRO16-4, size \qtyproduct{2.8x1x0.6}{\milli\metre}, mass \SI{7.98}{\milli\gram}), measured in two-probe configuration, is presented in \cref{fig:Simultaneous_MagRes}.
Sample \#2 (CR19-17, size \qtyproduct{2x1x0.1}{\milli\metre}, mass \SI{2.51}{\milli\gram}), measured in four-probe configuration with an additional thermometer connected directly to its top-surface, is presented in \cref{fig:Top_thermometer}.
Further technical details are given in the following sections.

\section{Results and discussion}
\subsection{Simultaneous electrical-transport and magnetic measurements}
\begin{figure}[tb]
\includegraphics[page=1,width=89mm]{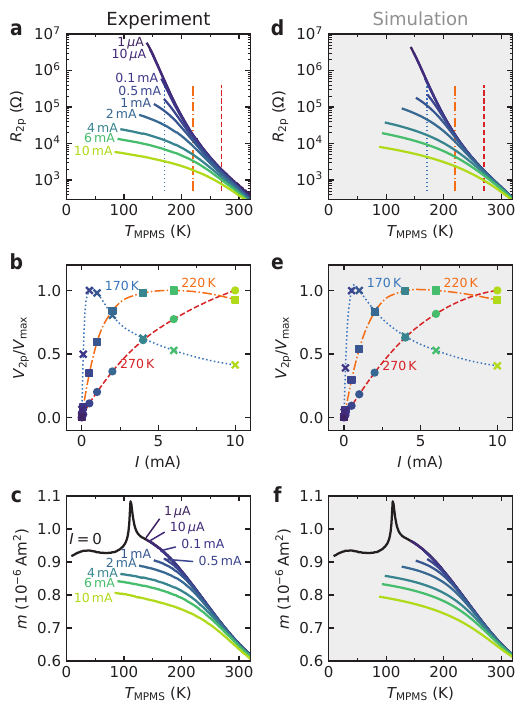}

\subfloat{\label{fig:Res}}
\subfloat{\label{fig:IVs}}
\subfloat{\label{fig:Mag}}
\subfloat{\label{fig:ResSim}}
\subfloat{\label{fig:IVsSim}}
\subfloat{\label{fig:MagSim}}
\caption{\textbf{Simultaneous magnetic and electrical transport measurements with constant current.}
		\protect\subref{fig:Res} Experimental sample resistance as a function of temperature for different applied currents.
		The resistance is measured in a two-probe configuration on Sample \#1.
		\protect\subref{fig:IVs} Voltage--current characteristics for selected temperatures as indicated by the vertical linecuts in \protect\subref{fig:Res}.
		The voltage curves are normalised to their maximum value $V_\mathrm{max}$.
		\protect\subref{fig:Mag} Sample magnetic moment measured simultaneously with the resistance.
		For magnetic measurements, the data for $I=0$ is also included.
		\protect\subref{fig:ResSim} Simulated data of sample resistance,
		\protect\subref{fig:IVsSim} voltage--current characteristics, and
		\protect\subref{fig:MagSim} magnetic moment
		 calculated by using the self-heating model of \cref{eq:JouleSelf} with $\alpha=\SI{180}{\kelvin\per\watt}$.
}
\label{fig:Simultaneous_MagRes}

\end{figure}

\begin{figure*}[tb]
\includegraphics[page=2,width=160mm]{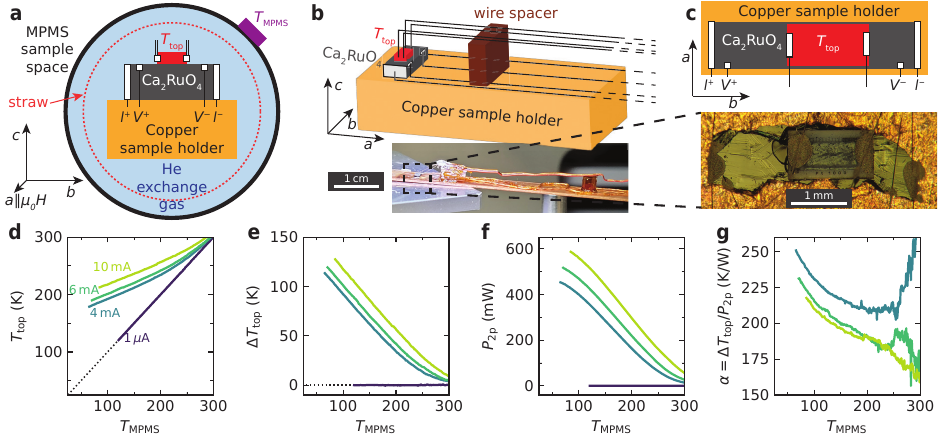}

\subfloat{\label{fig:Setup_MPMS}}
\subfloat{\label{fig:Setup_copper}}
\subfloat{\label{fig:Setup_sample}}
\subfloat{\label{fig:Ttop}}
\subfloat{\label{fig:dTtop}}
\subfloat{\label{fig:P2p}}
\subfloat{\label{fig:alpha}}
\caption{\textbf{Sample self-heating evaluated with a top thermometer.}
	\protect\subref{fig:Setup_MPMS} Schematics of the sample and holder inside the MPMS and location of the system thermometer $T_\mathrm{MPMS}$.
	\protect\subref{fig:Setup_copper} Schematic drawing and corresponding photograph of the sample holder with kapton wire spacer and
	\protect\subref{fig:Setup_sample} detail of the sample itself.
	The spatial directions are indicated with \CRO orthorhombic crystalline axes $a$, $b$, and $c$.
	\protect\subref{fig:Ttop} Top-thermometer temperature as a function of system temperature for several currents and 
	\protect\subref{fig:dTtop} corresponding increase of sample temperature due to current heating.
	\protect\subref{fig:P2p} Electrical power supplied to the sample by the flowing current and
	\protect\subref{fig:alpha} experimental values of $\alpha(T)$ if the temperature increase is described by the Joule self-heating model of \cref{eq:JouleSelf}.
}
\label{fig:Top_thermometer}

\end{figure*}

In \cref{fig:Res}, we show the resistance of Sample \#1 in a 2-probe configuration $\Rp$.
For the smallest current $I=\SI{1}{\micro\ampere}$, the resistance increases several orders of magnitude upon lowering temperature and it goes beyond the measurement limit at about \SI{150}{\kelvin}, consistent with previous reports of high-quality \CRO crystals \cite{
nakatsuji1997ca2ruo4,
nakatsuji2000quasi
}.
For larger values of current up to $I=\SI{10}{\milli\ampere}$, the resistance curves gradually become lower, in accordance with other reports \cite{
okazaki2013current,
cirillo2019emergence,
zhao2019nonequilibrium,
terasaki2020non
}.
By taking vertical linecuts in \cref{fig:Res} (dotted lines), we extract voltage--current characteristics at three fixed values of $\TMPMS$.
The resulting curves in \cref{fig:IVs} show a non-linear behaviour that becomes more pronounced at lower temperatures.
Similar trends have been previously reported for voltage--current characteristics of \CRO \cite{
tsurumaki2020stable,
avallone2021universal
}.
We note that the location of the region of negative differential resistance, after the voltage peak, is strongly dependent on the thermal couplings and the sample temperature, and it has been suggested to be the fingerprint of \CRO metal--insulator transition \cite{
	okazaki2013current,
	sakaki2013electric,	
	fursich2019raman,
	zhang2019nano
}.

As an important additional probe for \CRO properties under applied current, we also measure the sample magnetic moment $m$ simultaneously with the resistance.
We present in \cref{fig:Mag} the magnetic moment for $I=0$ which, upon decreasing temperature, shows a gradual increase, a peak at about \SI{110}{\kelvin} indicating \CRO antiferromagnetic transition, and a final saturating trend, consistent with literature \cite{
braden1998crystal
}.
Note that we intentionally report the measured magnetic moment $m$ instead of the sample magnetisation $M=m_\mathrm{sample}/V_\mathrm{sample}$ because $m$ may contain additional background signals as discussed in the following section.
With applied current $I>0$, the antiferromagnetic transition disappears and the magnetic moment decreases.
As for the resistance, the measurements are interrupted whenever it becomes impossible to source the chosen current to the sample.

In order to identify non-thermal current-induced effects, we attempt to estimate and subtract the Joule self-heating.
For this purpose, we consider a simple model in which the sample is at an effective temperature $\TMPMS + \Delta T$, and the $\Delta T$ is determined solely by the electrical power $P=I^2R$ as
	\begin{equation}
		\label{eq:JouleSelf}
		\Delta T (\TMPMS) = \alpha P(\TMPMS+\Delta T) = \alpha I^2 R_0 (\TMPMS+\Delta T),
	\end{equation}
where $\alpha$ is a constant expressing thermal resistance between the sample and the cryostat and $R_0$ is the sample resistance with close-to-zero current.
By using as only input $R_0(T) = \Rp(T, I{=}\SI{1}{\micro\ampere})$, we simulate the data in \cref{fig:ResSim,fig:IVsSim} by solving \cref{eq:JouleSelf}.
We also simulate the magnetisation data in \cref{fig:MagSim} using $m_0=m(I{=}0)$.
The data is produced by adjusting the value of the phenomenological constant $\alpha=\SI{180}{\kelvin\per\watt}$, for which we find a striking qualitative agreement between experiment and simulation.
The model correctly captures the current-induced reduction of both $\Rp$ and $m$, and also the non-linear trend of the current--voltage characteristics.
This indicates that a significant portion of the observed behaviour can be explained by Joule self-heating of a temperature-homogeneous insulating phase, underlying the importance of developing a special technique to accurately measure the sample temperature.

\subsection{Joule self-heating of \CRO}

In order to accurately assess the sample temperature, we perform another set of measurements on a similar \CRO crystal (Sample \#2) connected in a 4-probe configuration (full data in \cref{fig:RawSample2}).
Direct comparison of the two- and four-probe resistance indicates that the contact resistance is negligible (\cref{fig:Resistance24p}).
A thermometer glued by GE Varnish 7031 directly on the sample top surface is used to measure $\Ttop$ as shown in \cref{fig:Setup_MPMS}.
For this purpose, we chose a platinum resistive sensor (Heraeus Pt1000, SMD0603) whose substrate was thinned down to about \SI{80}{\micro\metre} thickness by mechanical polishing in order to enhance its proximity to the sample.
We also mechanically removed the sensor contact pads, which contain magnetic materials such as nickel, in order to bring its magnetic signal to a negligible value (\cref{fig:MagneticBG}).
We provided electrical contact to the sensor (\cref{fig:Setup_sample}) by using Ag paint and two thin Au wires (diameter \SI{18}{\micro\metre}, length \SI{0.5}{\centi\metre}) which are then connected in a four-probe configuration to a set of phosphor bronze wires that have low thermal conductivity.
As shown in \cref{fig:Setup_copper}, we minimise heat escape from the temperature sensor by using kapton spacers that keep its wires physically separated from the highly conductive sample holder.
Because the magnetic moment of Sample \#2 is rather small, we perform magnetic measurements with both a positive and negative applied current in order to identify the magnetic signal generated by the electrical leads (detailed description in \cref{fig:AmpereBG}).
This background signal is subtracted to extract the value of $m$ for this sample shown in \cref{fig:Universal_Rmu}.

In \cref{fig:Ttop}, we observe that $\Ttop$ significantly deviates from $\TMPMS$ (dotted line), indicating that the sample is substantially heated by the flowing current, especially at lower temperatures.
We quantify such sample heating in \cref{fig:dTtop} as $\dTtop = \Ttop - \TMPMS$ and also calculate in \cref{fig:P2p} the electrical power dissipated by the flowing current as $\Pp=I^2 \Rp$.
Since the power dissipation through the low-resistance copper leads is negligible, most of $\Pp$ is dissipated through the sample and at its electrical contacts.
To test whether the dissipated power determines a Joule self-heating in accordance with the model of \cref{eq:JouleSelf}, we calculate the experimental $\alpha_\mathrm{top}(T)=\dTtop/\Pp$ in \cref{fig:alpha}.
We find values $\alpha_\mathrm{top}=\SIrange{150}{250}{\kelvin\per\watt}$ which are consistent with the value $\alpha=\SI{180}{\kelvin\per\watt}$ used in the simulation of \cref{fig:Simultaneous_MagRes}, thus supporting our model choice.
The increase of $\alpha_\mathrm{top}$ at lower temperatures indicates a worse sample cooling, possibly due to a decreased thermal conductivity of the components or to a lower pressure of the He exchange gas.

\subsection{Universal relationship between magnetic moment and resistance}
\begin{figure}[tb]
\includegraphics[page=3,width=89mm]{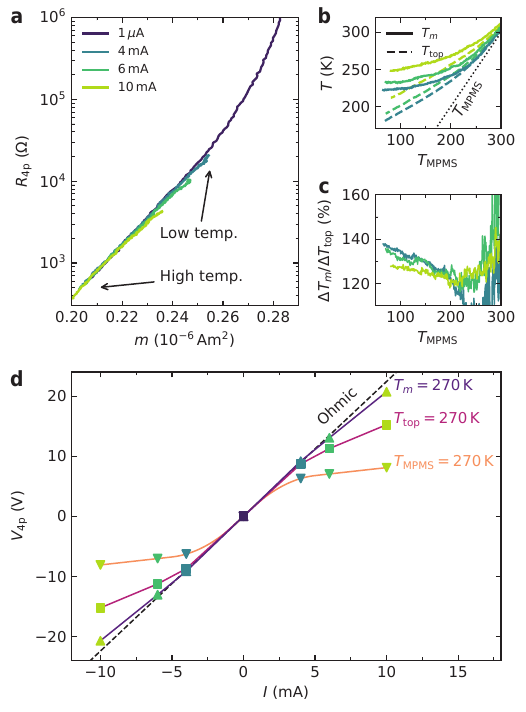}

\subfloat{\label{fig:Rmu}}
\subfloat{\label{fig:dTm}}
\subfloat{\label{fig:dTm_ratio}}
\subfloat{\label{fig:IVs_different}}
\caption{\textbf{Magnetic moment as internal thermometer.}
		\protect\subref{fig:Rmu} Relationship between the resistance and magnetic moment of Sample \#2 for different temperatures and currents.
		The colour indicates different applied currents, while the temperature dependence is implicit.
		\protect\subref{fig:dTm} Sample heating estimated from the magnetic moment $\Tm$, the top thermometer $\Ttop$, and
		\protect\subref{fig:dTm_ratio} their ratio.
		\protect\subref{fig:IVs_different}
		Voltage--current characteristics at a fixed sample temperature evaluated with different probes.
		As the probe of sample temperature becomes more accurate, the curves tend to a more ohmic behaviour.
}
\label{fig:Universal_Rmu}

\end{figure}

To reveal possible non-thermal current-induced effects that would induce different changes of resistance and magnetic moment, we plot in \cref{fig:Rmu} the data of the 4-probe resistance $\Rpp$ vs $m$ for Sample \#2.
The data mostly collapses on the same curve, suggesting a universal correlation between $\Rpp$ and $m$, irrespective of the applied current value.
This is a surprising result because if the band structure of \CRO is changed by the flowing current, there is no expectation that both $\Rpp$ and $m$ change in the same manner.
Despite the extensive overlap of the curves, some deviation is observed in the high-resistance high-magnetisation region, which corresponds to lower temperatures.

To investigate these deviations, we tentatively assume that $m$ does not depend on current but only on temperature and we use the experimental data of $m(\TMPMS)$ to calculate the sample ``magnetic temperature" $\Tm$.
Under our assumption, $\Tm$ provides an internal probe of sample temperature, which we use to estimate the sample heating as $\dTm = \Tm - \TMPMS$.
This heating is systematically larger than what is measured by $\Ttop$, and their ratio in \cref{fig:dTm_ratio} shows that $\dTm$ is up to \SI{40}{\percent} larger than $\dTtop$, implying that the top thermometer measures a value which is significantly lower than the sample average temperature.

We show in \cref{fig:IVs_different} the voltage--current characteristics for Sample \#2 extracted at a constant sample temperature of \SI{270}{\kelvin} estimated by different temperature probes.
Changing the temperature probe from $\TMPMS$, to $\Ttop$, to $\Tm$, the non-linear curves become more and more straight, and approach the ohmic behaviour.
This indicates that a large component of the observed non-linearity can be ascribed to an underestimation of the average sample temperature caused by the Joule self heating.
Non-thermal current-induced effects, if present, should be investigated after removing this large heating component.
We note that some deviation from the ohmic behaviour persists even when using $\Tm$, which may indicate the presence of non-thermal current-induced effects that will be further investigated in the following section.

\subsection{Possible sample temperature inhomogeneity}
\begin{figure}[tb]
\includegraphics[page=4,width=89mm]{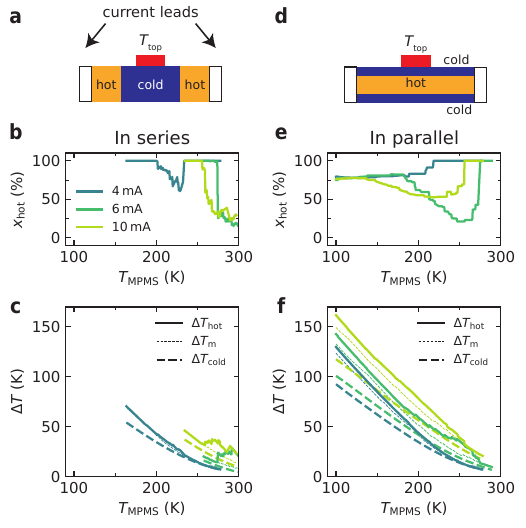}

\subfloat{\label{fig:Series}}
\subfloat{\label{fig:perc_Series}}
\subfloat{\label{fig:T_Series}}
\subfloat{\label{fig:Parallel}}
\subfloat{\label{fig:perc_Parallel}}
\subfloat{\label{fig:T_Parallel}}

\caption{\textbf{Possible spatial temperature inhomogeneity.}
		\protect\subref{fig:Series} Schematic of a simplified sample temperature inhomogeneity with vertical (i.e., in series) and
		\protect\subref{fig:Parallel} horizontal (i.e., in parallel) boundaries.
		In both cases, we set the temperature of the colder regions to $T_\mathrm{cold} = \Ttop$.
		\protect\subref{fig:perc_Series} Simulated hot volume fraction and
		\protect\subref{fig:T_Series} temperature of the hotter region for the series configuration, and
		\protect\subref{fig:perc_Parallel},
		\protect\subref{fig:T_Parallel} for the parallel configuration.
		Also here, all the temperature increases $\Delta T$ are referred to $\TMPMS$.
}
\label{fig:IV_Thermometer}

\end{figure}

We now discuss whether residual deviations of the $R$ vs $m$ curves can be described by possible inhomogeneities of the sample temperature.
We note that the magnetic moment is a bulk measurement averaged over the entire sample volume, while the resistance is dominated by the most-conductive electrical channel.
To account for possible inhomogeneities, we consider two simplified model scenarios in which the sample temperature presents hotter regions with vertical boundaries, which we call \textit{in series} (\cref{fig:Series}), or horizontal, which we call \textit{in parallel} (\cref{fig:Parallel}).
The first scenario can be related to excess sample heating in proximity of the current leads, possibly due to contact resistance.
The second scenario can be related to an excess heating in the internal part of the sample, that is further away from the colder bottom and top surfaces that are in contact with the sample holder and exchange gas, respectively.
The formulation of the following analysis allows the location and extent of the hotter and colder regions to be different from the one in the schematic drawings, as long as the directionality is respected (for example, the hotter regions could be multiple or spatially asymmetric).

In this simplified model, we consider two sharply defined regions at a hotter ($T_\mathrm{hot}$) and colder ($T_\mathrm{cold}$) temperature whose extent is defined by the volume fraction $x_\mathrm{hot}$.
For both scenarios, the sample magnetisation is given by the volume average
\begin{equation}
	\label{eq:mag}
	m(T) = \xh m_0(\Th) + (1-\xh) m_0(T_\mathrm{cold}),
\end{equation}
where $m_0$ is the magnetic moment at zero current.
The sample resistance, instead, is calculated differently in the two scenarios as
\begin{equation}
	\label{eq:res}
	\begin{aligned}
	R^\mathrm{series}(T) =& \xh R_0(\Th) + (1-\xh) R_0(T_\mathrm{cold}),
	\\
	\frac{1}{R^\mathrm{parallel}(T)} =& \frac{\xh}{ R_0(\Th)} + \frac{(1-\xh)}{ R_0(T_\mathrm{cold})}.
	\end{aligned}
\end{equation}
Following the discussion of the previous section, we expect the sample average temperature to be on average larger than what is measured by $\Ttop$.
We thus tentatively set $T_\mathrm{cold}=\Ttop$ and use the experimental data of $m(T)$ and $R(T)$ as inputs to solve the coupled set of equations \cref{eq:mag,eq:res} to find numerical solutions for $\xh$ and $\Th$.

In the series scenario of \cref{fig:perc_Series,fig:T_Series}, the values of $\xh$ are not well defined in the higher-temperature region because the sample self-heating is small (i.e., $\Th\sim T_\mathrm{cold}$).
In this region, the universality of resistance vs magnetisation is satisfied (\cref{fig:Rmu}).
At lower temperatures, $x_\mathrm{hot}$ shoots up to \SI{100}{\percent}, indicating that a large portion of the sample is at $T_\mathrm{hot}$.
For larger applied currents, no numerical solution is found below about \SI{200}{\kelvin}, indicating that the series temperature inhomogeneity cannot explain the experimental behaviour.
The formation of a hotter regions with vertical boundary is thus unlikely, indicating that sample self-heating at the current leads is negligible, consistent with our estimate of a low contact resistance (\cref{fig:Resistance24p}).

In the parallel scenario of \cref{fig:perc_Parallel}, a numerical solution is found for all temperatures and currents.
At low temperature, $x_\mathrm{hot}$ approaches a value of about \SI{80}{\percent} that is consistent for all experimental currents, indicating that the coexistence of a broad hotter region and a thin colder region is a possible description of the experimental behaviour.
From \cref{fig:T_Parallel}, we note that the temperature difference between the hotter and colder regions $\Delta\widetilde{T}=\Delta T_\mathrm{hot}- \Delta T_\mathrm{cold}$ is of a few Kelvin at room temperature (electrical power $\Pp\sim\SI{50}{\milli\watt}$ from \cref{fig:P2p}), while it grows to about $\Delta\widetilde{T} \sim\SI{30}{\kelvin}$ ($\Pp\sim\SI{500}{\milli\watt}$) at low temperature.
Considering that thermal conduction within the sample is given by $P=\kappa\dfrac{A}{t}\Delta \widetilde{T}$, where the room-temperature conductivity is $\kappa_{c, \CRO}=\SI{1.8}{\watt\per\meter\per\kelvin}$ \cite{
kawasaki2021thermal
} and $A/t$ is the sample cross-sectional area over its thickness, we estimate that at room temperature $\Delta \widetilde{T}\sim \SI{1.4}{\kelvin}$ along the $c$ direction of \CRO.
At lower temperatures, this vertical temperature inhomogeneity grows up to a factor 10 due to the increasing electrical power, and may be further enhanced by the decreasing thermal conductivity of \CRO.
The presence of hot and cold regions in parallel is thus a reasonable possibility, and their extent may depend on sample size, thermal couplings, and cooling conditions.

\section{Conclusions}
We have investigated current-induced phenomena in \CRO through a wide temperature range by means of simultaneous magnetic and electrical measurements.
Despite the purpose-made setup, the sample experienced a large Joule self-heating that we quantified by means of a simple model and a thermometer in direct contact with the sample.
While most deviations from ohmic behaviour can be explained by homogeneous sample heating, additional effects are present.
Temperature inhomogeneity is intrinsic to a current-induced steady state where the continuous heat input is balanced by the heat escape.
Therefore, we introduced a model of inhomogeneous sample heating which explained most of these additional non-linearities as due to a temperature gradient in the direction perpendicular to the current flow.
This analysis allowed us to identify that a combination of homogeneous and inhomogeneous current-induced heating are responsible for the observed behaviour.
Non-thermal current-induced effects in \CRO, if present, are below the detection limit of our experiment.
Our results pose a solid basis for investigating current-induced phenomena in insulators, where large current heating is unavoidable.
\section*{Acknowledgements}
The authors thank N. Manca for valuable comments on the manuscript and H. Michishita for machining the copper components.
This work was supported by JSPS Grant-in-Aids KAKENHI Nos. JP26247060, JP15H05852, JP15K21717, JP17H06136, and JP18K04715, as well as by JSPS Core-to-Core program.
G.M. acknowledges support from the Dutch Research Council (NWO) through a Rubicon grant number 019.183EN.031, and support from the Kyoto University Foundation.


\bibliography{
	Biblio_Magnetism,
	Biblio_NonEquilibrium,
	Biblio_QMaterials,
	Biblio_Ruthenates
}
%

	\onecolumngrid
\appendix
\newpage
\noindent\rule{1\columnwidth}{1pt}
\centerline{\huge\texttt{Supplementary Material}}\\
\noindent\rule{1\columnwidth}{1pt}

\renewcommand\thefigure{S\arabic{figure}}    
\setcounter{figure}{0}
\renewcommand\thetable{S\arabic{table}}    
\setcounter{table}{0}
\renewcommand\theequation{S\arabic{equation}}    
\setcounter{equation}{0}


\begin{figure}[h]
	\includegraphics[page=5,width=80mm]{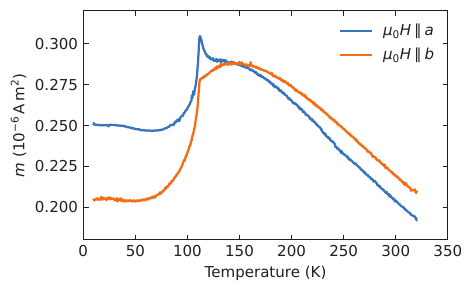}
	
	\caption{
		\textbf{Determination of in-plane crystalline axes.}
		Before mounting on the copper sample holder, the magnetic signal of Sample \#2 alone is measured in a standard MPMS straw in two different sample orientations.
		The absence of the peak at \SI{110}{\kelvin} in the orange curve allows us to identify the crystalline orthorhombic $b$ axis which is the easy axis for antiferromagnetic ordering in \CRO.
	}
	
	\label{fig:SampleAxes}
	
\end{figure}

\begin{figure}[h]
	\includegraphics[page=6,width=89mm]{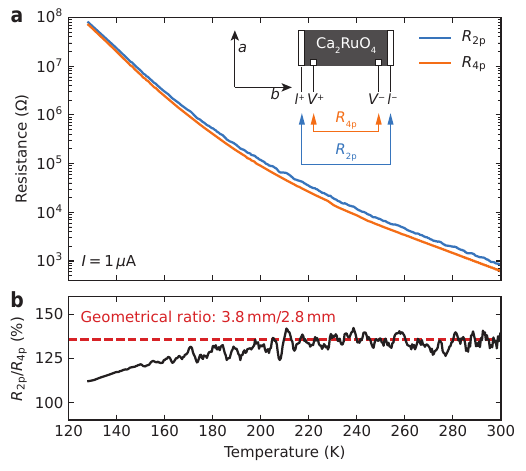}
	
	\subfloat{\label{fig:Resistance24p_curves}}
	\subfloat{\label{fig:Resistance24p_ratio}}
	
	\caption{
		\textbf{Comparison of two- and four-probe resistance for Sample \#2.}
		\protect\subref{fig:Resistance24p_curves} Resistance vs temperature curves measured with a current of \SI{1}{\micro\ampere} and	\protect\subref{fig:Resistance24p_ratio} their ratio.
		At higher temperature, the ratio $\Rp/\Rpp$ is compatible with the geometrical ratio of the sample width \SI{3.8}{\milli\metre} and the inner distance between the voltage probes \SI{2.8}{\milli\metre}.
		This indicates that $\Rp$ is larger merely due to geometry (i.e., due to a longer electrical channel), thus indicating that contact resistance is negligible.
		The data indicates that the contact resistance is smaller than \SI{100}{\ohm} at room temperature and it does not increase significantly at lower temperatures.
		At lower temperature, instead, the ratio significantly deviates from the geometrical one.
		This devtiation cannot be explained as to be due to contact resistance, which would determine an increase of $\Rp$, but it is possibly related to the formation of temperature inhomogeneity within the sample as discussed in \cref{fig:IV_Thermometer} of the main text.
	}
	
	\label{fig:Resistance24p}
	
\end{figure}

\begin{figure}[h]
	\includegraphics[page=7,width=89mm]{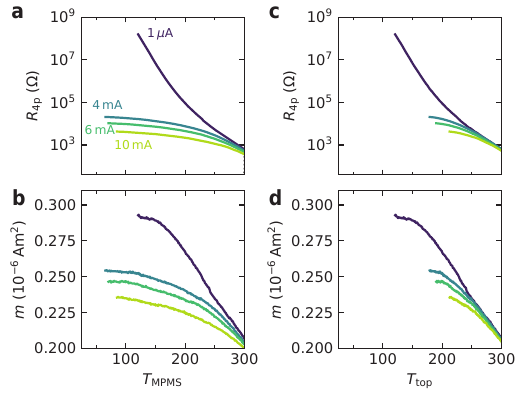}
	
	\subfloat{\label{fig:S2_RvsT}}
	\subfloat{\label{fig:S2_mvsT}}
	\subfloat{\label{fig:S2_RvsTtop}}
	\subfloat{\label{fig:S2_mvsTtop}}	
	
	\caption{
		\textbf{Resistance and magnetic curves as a function of temperature for Sample \#2.}
		\protect\subref{fig:S2_RvsT} Four-probe resistance and
		\protect\subref{fig:S2_mvsT} magnetic moment measured for different values of current similar to what presented for Sample \#1 in \cref{fig:Simultaneous_MagRes} of the main text.
		The data is plotted as a function of the system thermometer $\TMPMS$.
		\protect\subref{fig:S2_RvsTtop} Resistance and
		\protect\subref{fig:S2_mvsTtop} magnetic moment replotted as a function of the top thermometer $\Ttop$.
		Even if the effects of current look smaller in the data with $\Ttop$, the current-induced reduction of $\Rpp$ and $m$ is qualitatively the same.
		This data is used for the analysis of \cref{fig:Top_thermometer,fig:Universal_Rmu,fig:IV_Thermometer} in the main text.
	}
	
	\label{fig:RawSample2}
	
\end{figure}

\begin{figure}[h]
	\includegraphics[page=8,width=80mm]{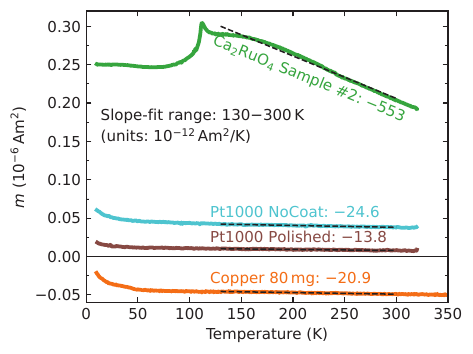}
	
	\caption{
		\textbf{Comparison of magnetic backgrounds.}
		Magnetic signals of isolated materials measured in the MPMS.
		(1) \CRO Sample \#2;
		(2) Pt1000 sensor after mechanical removal of the electrode coating and
		(3) same Pt1000 sensor after also thinning the substrate down to about \SI{80}{\micro\metre} by mechanical polishing;
		(4) A small cubic copper piece of about \SI{80}{\milli\gram}.
		When compared to \CRO, the magnetic signal of the polished Pt1000 temperature sensor is negligible.
		The copper material has a considerable magnetic moment, but the signal of the long copper stripe used as sample holder does not appear in the magnetic measurements performed in this work thanks to its homogeneous shape.
		This is because the MPMS employs a SQUID magnetometer that is only sensitive to spatial variations of magnetic moment.
		Fits in the range \SIrange{130}{300}{\kelvin} show that the magnetic temperature coefficient of \CRO is more than one order of magnitude larger than the other materials.
		This indicates that the insurgence of magnetic signals due to local heating of the background materials, which is proportional to their temperature coefficient, is minimal in our setup [G. Mattoni \textit{et al.}, APL \textbf{116} (2020)].
	}
	
	\label{fig:MagneticBG}
	
\end{figure}

\begin{figure*}[h]
	\includegraphics[page=9,width=150mm]{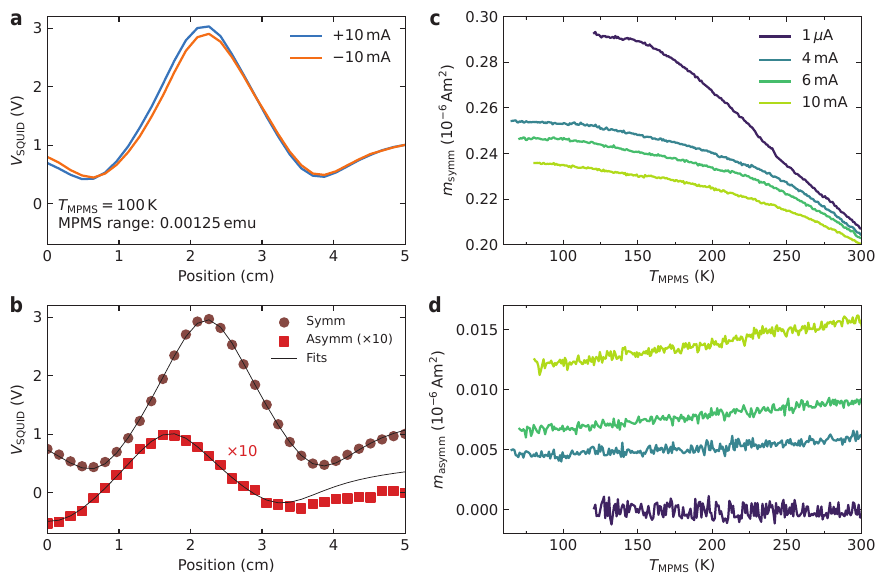}
	
	\subfloat{\label{fig:VSQUID_pm}}
	\subfloat{\label{fig:VSQUID_symm}}
	\subfloat{\label{fig:mSymm}}
	\subfloat{\label{fig:mAsymm}}
	
	\caption{
		\textbf{Removal of magnetic signal generated by the current leads.}
		In order to identify and subtract the magnetic signal coming from the current leads, each magnetic measurement is repeated with opposite values of current.
		\protect\subref{fig:VSQUID_pm} SQUID response measured with a positive and negative current of \SI{10}{\milli\ampere}.
		\protect\subref{fig:VSQUID_symm} Symmetric [$(V_\mathrm{SQUID}(I^+) + V_\mathrm{SQUID}(I^-))/2$] and asymmetric [$(V_\mathrm{SQUID}(I^+) - V_\mathrm{SQUID}(I^-))/2$] components of the SQUID signal.
		The latter is amplified by a factor 10 for clarity.
		Both data are fitted with the SQUID response function to extract the corresponding magnetic moment (for details see [G. Mattoni \textit{et al.}, APL \textbf{116} (2020)]).
		The fit for the symmetric component has good agreement with the data and provides a magnetic signal which is centred at position $x=\SI{2.2}{\centi\metre}$, consistent wtih the sample spatial location.
		The fit for the asymmetric component, instead, reveals a magnetic signal located at a shifted position $x=\SI{1.8}{\centi\metre}$ which is not produced by the sample but from the current flowing through the current leads.
		\protect\subref{fig:mSymm} Magnetic signal of Sample \#2 that is used in the manuscript and
		\protect\subref{fig:mAsymm} spurious magnetic signal of the current leads that is correctly discarded thanks to the described measurement technique.
	}
	
	\label{fig:AmpereBG}
	
\end{figure*}

\end{document}